# Moiré synaptic transistor for homogeneous-architecture reservoir computing

Pengfei Wang(王鹏飞)[1,†], Moyu Chen(陈墨雨)[1,†], Yongqin Xie(谢永勤)[1], Chen Pan(潘晨)[2], Kenji Watanabe[3], Takashi Taniguchi[4], Bin Cheng(程斌)[2,\*], Shi-Jun Liang(梁世军)[1,\*], and Feng Miao(缪峰)[1,\*]

[1] Institute of Brain-Inspired Intelligence, National Laboratory of Solid State Microstructures, School of Physics, Collaborative Innovation Center of Advanced Microstructures, Nanjing University, Nanjing 210093, China.
[2] Institute of Interdisciplinary Physical Sciences, School of Science, Nanjing University of Science and Technology, Nanjing 210094, China.
[3] Research Center for Functional Materials, National Institute for Materials Science, 1-1 Namiki, Tsukuba 305-0044, Japan.
[4] International Center for Materials Nanoarchitectonics, National Institute for Materials Science, 1-1 Namiki, Tsukuba 305-0044, Japan.

† These authors contributed equally to this work.
\*Corresponding authors. Email: bincheng@njust.edu.cn; sjliang@nju.edu.cn; miao@nju.edu.cn

Reservoir computing has been considered as a promising intelligent computing paradigm for effectively processing complex temporal information. Exploiting tunable and reproducible dynamics in the single electronic device have been desired to implement the "reservoir" and the "readout" layer of reservoir computing system. Two-dimensional moiré material, with an artificial lattice constant many times larger than the atomic length scale, is one type of most studied artificial quantum materials in community of material science and condensed-matter physics over the past years. These materials are featured with gate-tunable periodic potential and electronic correlation, thus varying the electric field allows the electrons in the moiré potential per unit cell to exhibit distinct and reproducible dynamics, showing great promise in robust reservoir computing. Here, we report that a moiré synaptic transistor can be used to implement the reservoir computing system with a homogeneous reservoir-readout architecture. The synaptic transistor is fabricated based on a h-BN/bilayer graphene/h-BN moiré heterostructure, exhibiting ferroelectricity-like hysteretic gate voltage dependence of resistance. Varying the magnitude of the gate voltage enables the moiré transistor to be switched between long-term memory and short-term memory with nonlinear dynamics. By employing the short- and long-term memory as the reservoir nodes and weights of the readout layer, respectively, we construct a full-moiré physical neural network and demonstrate that the classification accuracy of 90.8% can be achieved for the MNIST handwritten digit database.

Our work would pave the way towards the development of neuromorphic computing based on the moiré materials.

## INTRODUCTION

Reservoir computing, as one of important neuromorphic computing algorithms, has received much attention over the past years, due to its powerful capability in time series processing[1-3]. The architecture of reservoir computing includes "reservoir" layer and "readout" layer. Recently, there have been many reports focusing on the hardware implementation of reservoir computing with distinct neuromorphic material systems[4], such as memristive materials[5-9], two-dimensional semiconducting materials[10-13], nanowire materials[14, 15], ferroelectric materials [16, 17], ferromagnetic materials[18-20], polymer semiconductor[21-23], and so on. Most of these works focus on the implementation of "reservoir" layer by emulating the synaptic plasticity, which governs the modulation of connection strengths between neurons and underpins learning, cognition and memory processes in the human brain[24]. Most recently, all-ferroelectric or memristive hardware implementation of reservoir computing has been proposed[25, 26]. However, the operating mechanisms of the synaptic devices are based on field-driven ion motion, and the inherent stochastic dynamics would impose a fundamental limitation to the device uniformity of large-scale integrated arrays[27-31]. Thereby, it remains a pressing need to explore new materials to design synaptic devices, in which tunable and reproducible electron dynamics can be used to implement the "reservoir" and the "readout" layer of reservoir computing.

Moiré systems have latterly garnered significant attention owing to their exotic properties and extraordinary tunability. In moiré heterostructures, the interference between lattices of adjacent two layers creates the moiré patterns, resulting in novel emergent phenomena including strong electron correlation[32-39], unconventional superconductivity[34, 40-45], interfacial ferroelectricity[46-54], etc. These quantum states in moiré systems can be finely tuned by manipulating accessible external stimuli such as electrical fields, light, or strain, thus creating a novel highly tunable platform that will undoubtedly expedite the advance of future computing, particularly in the realm of neuromorphic computing[55, 56]. In this work, we propose a moiré synaptic transistor by encapsulating Bernal-stacked bilayer graphene into two h-BN layers. This moiré synaptic device exhibits a unique and tunable memory behavior, capable of faithfully emulating both short-term plasticity (STP) and long-term plasticity (LTP). The short-term memory of the device enables the implementation of reservoir layer for processing temporal information, while the stable long-term memory characteristics allow for implementing readout layer to achieve continuous modulation of the synaptic weight. Leveraging the high tunability of the moiré synaptic transistor, we demonstrate a full-moiré physical neural network (MPNN) and demonstrate that the recognition accuracy of the handwritten MNIST dataset can reach 90.8%. Our study opens up a new avenue for developing reservoir computing hardware based on moiré materials.

**RESULTS**

**Moiré synaptic transistor based on h-BN/BLG/h-BN heterostructure**

Reservoir computing represents a special kind of neuromorphic neural network. In this network, synapse plays a key role (Fig. 1a, bottom). Depending on the strength of the input stimuli, the synapse exhibits distinct plasticity: short-term plasticity (STP) and long-term plasticity (LTP). STP can be triggered by weak stimuli and occurs over milliseconds to minutes, enabling the processing of temporal information. In contrast, LTP can be induced by strong stimuli and lasts for hours or longer, responsible for learning and memory[24]. We design a moiré heterostructure synaptic transistor featuring electrically tunable electron polarization to faithfully emulate the distinct plasticity of the biological synapse (Fig. 1b). Due to the small mismatch (~1.8%) between h-BN and graphene lattices, a moiré pattern can be formed at the interface between the top h-BN and graphene by aligning these two layers (Fig. 1b). Notably, such moiré pattern gives rise to an interfacial moiré potential that could drastically alter electronic properties of the graphene. This moiré potential seen by graphene carriers can be tuned by applying a vertical electric field through gate stimulus, leading to exotic quantum phenomena. One particular quantum phenomenon is the interfacial ferroelectricity residing in moiré heterostructures comprised by Bernal bilayer graphene (BLG) and h-BN layers[46, 50]. In this ferroelectric system, the two distinct layers of graphene sense different strengths of moiré potentials, making the different vertical electric fields generated by gate stimuli have quite different capabilities of trapping the layer-polarized electrons in the ferroelectric switching process. Specifically, when a gate stimulus smaller than the threshold voltage is applied, the transferred electrons can only sense shallow moiré potentials, and thus are hard to be trapped, leading to a short-term response. On the contrary, a strong gate stimulus larger than the threshold voltage could substantially alter the moiré potential landscape sensed by the transferred layer-polarized electrons, facilitating the ferroelectric switch in a nonvolatile way. In this sense, a long-term response is expected.

We fabricate the moiré synaptic transistor by encapsulating a BLG into two h-BN thin flakes, with their long straight edges deliberately aligned, which leads to the coincidence of the crystallographic orientations of the h-BN layers and BLG. The assembled h-BN/BLG/h-BN stack is eventually deposited on a $SiO_2$/p-doped Si substrate before an Au top gate is fabricated (see Supplementary materials for fabrication details). Figures 1c and 1d show the schematic and the corresponding optical image of a typical moiré synaptic transistor, respectively, alongside the measurement setup. The top gate and the channel act as the pre-neuron and post-neuron terminals, respectively. The applied gate voltage ($V_G$) simulates the transmission of spikes from the axon of the pre-neuron to the post-neuron, and the measured drain-source current ($I_{DS}$) corresponds to the changes in the postsynaptic current (PSC).

We first carry out the measurements of transfer characteristics of the moiré synaptic transistor by varying the scanning range at room temperature, with the corresponding results shown in Fig. 1e and 1f. The data of $I_{DS}$ are acquired by sweeping $V_G$ back and forth between negative and positive values, at drain-source voltage ($V_{DS}$) of 0.1 V. Figure 1e illustrates the volatile transfer curve for a small $V_G$ sweeping range of -4 V to +4 V. In this curve, no significant hysteresis is observed, since the weak perpendicular electrical field cannot switch the ferroelectric polarization. When the $V_G$ sweeping range is expanded from -12 V to +12 V, a clear hysteretic behavior appears, leading to a memory window $\Delta V$ of 0.57 V, as illustrated in Fig. 1f. We further examine the memory window $\Delta V$ with various sweeping ranges (Fig. 1g), and find that $\Delta V$ is increased with the increase of the maximum value of the sweeping voltage ($V_{sweep, max}$) after $V_{sweep, max}$ is larger than a threshold value (~5V). The existence of such threshold voltage of the gate is attributed to the layer-specific moiré potential at the graphene/h-BN interface, which gives rise to layer-specific electronic correlation and anomalous electric screening effect[57, 58]. Our results indicate a transition from the volatile to the nonvolatile memory behavior of the device at the threshold gate voltage, suggesting that moiré synaptic transistor with tunable memory characteristics holds great promise for faithfully emulating synaptic plasticity including STP and LTP.

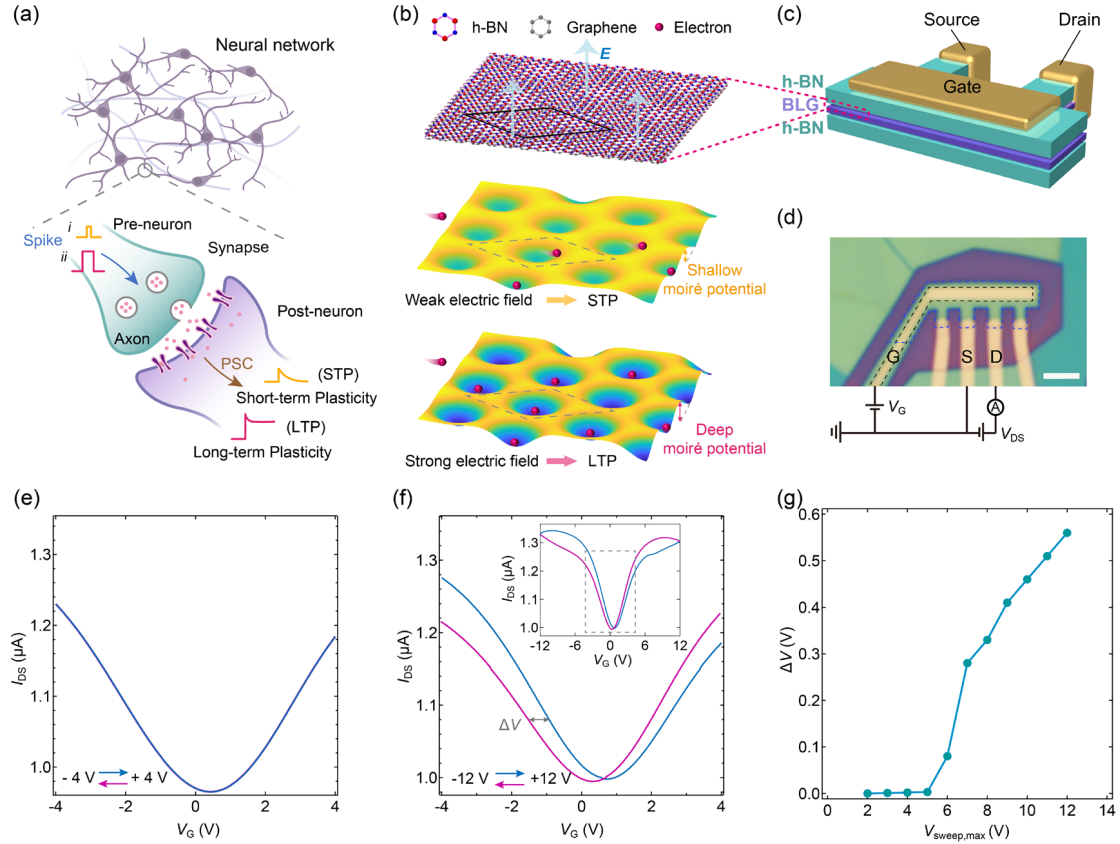

**Fig. 1. Design of moiré synaptic transistor.** (**a**) Schematic illustration of a biological neural system. Top, vast and intricate biological neural networks in the human brain. Bottom, schematic of a typical biological synapse. Weak stimuli (orange signal) and strong stimuli (red signal) induce PSCs of different intensities, respectively, showing the behavior transition of STP and LTP. (**b**) Schematics of the moiré superlattice and moiré potential near the h-BN/graphene interface. The moiré unit cell is highlighted by a black diamond. A vertical electric field (light blue arrow) is applied to the graphene electrons. Middle, when a weak electric field is present, the electrons sense a shallow moiré potential, resulting in a short-term unstable modulation to the layer-specified electron distribution. Bottom, when a strong electric field is present, the deep moiré potential strongly localizes electrons in the graphene upper layer, leading to long-term modulation. (**c**) Schematic of a typical moiré synaptic transistor structure based on van der Waals (vdW) moiré heterostructure. The BLG is aligned with the top h-BN crystal (red dashed box). (**d**) Optical microscope image and measurement setup of a moiré synaptic transistor. Scale bar, 4 μm. The top gate (G) and moiré heterostructure are highlighted by black and blue dashed boxes, respectively. S, source; D, drain. (**e**) Transfer characteristics ($I_{DS}$ versus $V_G$) of the moiré synaptic transistor at room temperature when $V_G$ sweeps between -4 V and 4 V, showing negligible hysteresis. (**f**) Plot of the enlargement for the gray dashed box in the inset that displays transfer characteristics at room temperature of the moiré synaptic transistor when $V_G$ sweeps between -12 V and +12V, exhibiting an obvious hysteresis. The shift of the curve reveals the memory window Δ$V$. The forward and backward scans are shown in blue and red, respectively. (**g**) Δ$V$ as a function of gate voltage sweep range ($V_{sweep, max}$).

**Emulation of short-term synaptic plasticity characteristic**

We investigate the dynamic response of the moiré synaptic transistor to input temporal signals. Figure 2a shows the measured current response at $V_{DS}$ = 0.1 V after applying a voltage pulse of 16 V for 4 s to the gate terminal. The current is significantly increased from 26.8 μA to 33.5 μA upon the application of the gate pulse, which is reminiscent of the excitatory postsynaptic current (EPSC) in biological synapses[59]. Following the termination of the gate pulse, the current does not immediately return to its initial state, instead, it drops to 27.2 μA and gradually decays back to the initial state within 100 s in an exponential decay manner. This behavior indicates a volatile memory characteristic and allows for emulation of STP. Such behavior stems from the fact that the ferroelectric polarization change induced by a weak electrical stimulus is inherently unstable and the resulting spontaneous polarization is depolarized within a short period.

In addition to emulation of STP, our transistor can generate the dynamic able to emulate the paired-pulse facilitation (PPF), which is characterized by a stepwise enhancement of the excitatory postsynaptic current (EPSC) magnitude upon repetitive stimulation and is usually involved in temporal information decoding[59]. Figure 2b shows the current response at $V_{DS}$ = 0.1 V after a pair of voltage pulses of 16 V for 4 s with an interval (Δ$t$) of 6 s is applied to the gate terminal. The resulting second current change ($A_2$) of ~ 0.47 μA is obviously larger than the first current change ($A_1$) of ~ 0.32 μA, resembling the behavior of PPF. We then present the PPF index ($A_2/A_1$) as a function of Δ$t$ to evaluate the accumulative effect, and find that the PPF index gradually decreases with increasing Δ$t$, as shown in Fig. 2c. The experimental results can be well fitted with a double exponential decay function validated by biological synapse[59]. This time-interval-dependent response of the device results from the fact that the ferroelectric polarization triggered by the second pulse continues to strengthen previous pulse-induced polarization when the pulse interval is shorter than the depolarization time.

We further explore the dynamic response of the moiré synaptic transistor to various external stimuli, *i.e.,* different pulse amplitude (Fig. 2d), different pulse width (Fig. 2e), and different pulse number (Fig. 2f). The excitatory postsynaptic current changes (ΔEPSC) are recorded at a fixed $V_{DS}$ of 0.1 V throughout the measurements. Figure 2d shows the ΔEPSC responses under a series of $V_G$ pulses with different amplitudes *(i.e.,* from 10 V to 18 V) but a fixed width of 3 s. The results show that the peak of ΔEPSC exhibits a substantial increment from ~ 6.6 µA to ~ 8.2 µA, as pulse amplitude increases. After removal of the pulse, the retained ΔEPSC also shows an increase from ~0.1 µA to ~ 0.8 µA, and takes a longer time to decay back to the initial state. This phenomenon can be attributed to the enhanced ferroelectric switching of polarization resulting from the increased pulse amplitude. In addition to the pulse amplitude, we vary the pulse width from 1 s to 9 s and monitor the ΔEPSC at a fixed pulse amplitude of 16 V, with results presented in Fig. 2e. Since the extended pulse width facilitates the ferroelectric polarization switching, it becomes evident that the retained current after pulse removal increases from ~ 0.18 µA to ~ 0.57 µA, accompanied by a longer relaxation time. Subsequently, by gradually increasing the number of applied pulses (16 V, 4 s, with a fixed pulse interval of 4 s) from 1 to 5, a significant cumulative effect is observed (as shown in Fig. 2f), which is manifested itself in the nonlinear increase of the ΔEPSC and relaxation time. These results demonstrate that the intrinsic short-term memory characteristics and nonlinearity of our device give rise to the dynamic changes of EPSC. These tunable and reproducible temporal dynamics allow for the clear distinction of different electrical stimuli and can be harnessed for processing complex temporal information.

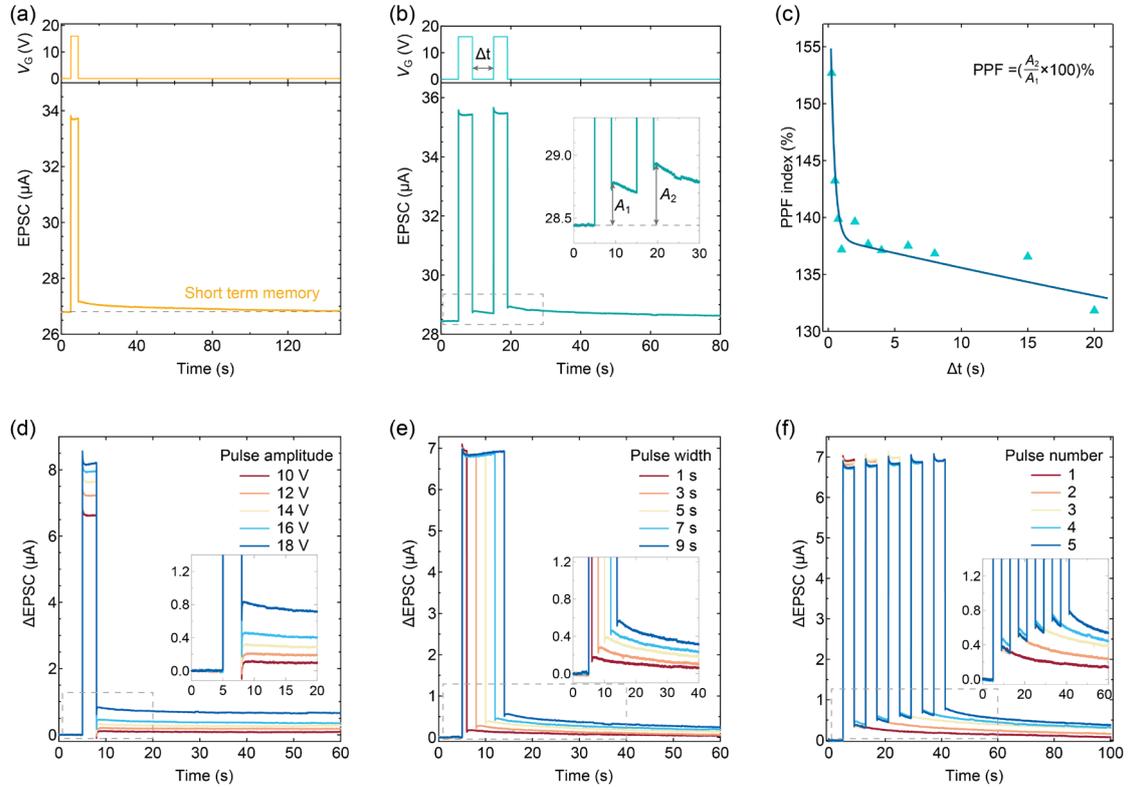

**Fig. 2. Characterization of short-term synaptic dynamics of the moiré synaptic transistor. (a)**

Current response induced by a $V_G$ pulse of 16 V for 4 s (at $V_{DS}$ = 0.1 V), indicating EPSC characteristic and STM effect. (**b**) PPF phenomenon, where the current response is induced by a pair of $V_G$ pulses (16 V, 4 s, with an interval ($\Delta t$) of 6 s). Inset, plot of the enlargement for the gray dashed box. (**c**) PPF index as a function of $\Delta t$. (**d** to **f**) Dynamic current change in response to various external stimuli by increasing the $V_G$ pulse amplitude from 10 V to 18 V (width, 4 s) (d), pulse width from 1 s to 9 s (amplitude, 16 V) (e), and pulse number from 1 to 5 (16 V, 4 s, with an interval of 4 s) (f). The inset in each figure is a magnified view of the dashed gray box in the current figure.

**Emulation of long-term synaptic plasticity characteristic**

We next proceed to study the long-term memory behavior of our moiré synaptic transistor. By further strengthening the external stimuli, the EPSC cannot recover to the pristine state for a long period, indicating a transition from STP to LTP (see Fig 2d-f and Fig. S2). We then apply a large positive $V_G$ pulse of 18 V for 4 s to the transistor and measure the current response of the transistor, with the results shown in Fig. 3a. We observe that the current is significantly increased by 1 μA compared to the initial state, and exhibits no noticeable degradation within 50 s, giving rise to a stable memory state. This nonvolatile conductance change effectively emulates long-term potentiation, which is associated with long-term memory (LTM). In contrast, the application of a relatively large $V_G$ pulse of -14 V for 4 s into the transistor results in a current decrease of 1 μA after pulse removal, as shown in Fig. 3b. The current experiences an exceedingly slow decay and eventually maintains a stable memory state, emulating the long-term depression (LTD).

The moiré synaptic transistor can exhibit multiple stable memory states through a pulse writing scheme. As shown in Fig. 3c, as a positive pulse sequence (with different pulse amplitudes but a fixed pulse width of 4 s) is applied to the device, the conductance of the device gradually and linearly increases from 280 μS to 330 μS. Notably, when we apply a sequence of negative pulses with different amplitudes but a fixed width of 4 s into the device, the conductance of the device gradually returns to its initial state of 280 μS, exhibiting a symmetric modulation behavior with respect to the positive pulses. Such capability of incrementally modulating the device's nonvolatile conductance serves as a fundamental requirement for achieving precise regulation of synaptic weights. Moreover, our moiré transistor exhibits a linear current-voltage characteristic over different conductance states (Fig. S3) and demonstrates excellent reproducibility with negligible cycle-to-cycle variation when subjected to repeated pulse sequences (Fig. S4). We also select eight conductance states and measure the retention time for each of these states, with results presented in Fig. 3d. We see that the conductance states are well separated and show no obvious degradation for up to $10^3$ s. These results indicate that our device can effectively emulate LTP and that its stable memory modulation characteristics hold great promise for the implementation of reservoir computing.

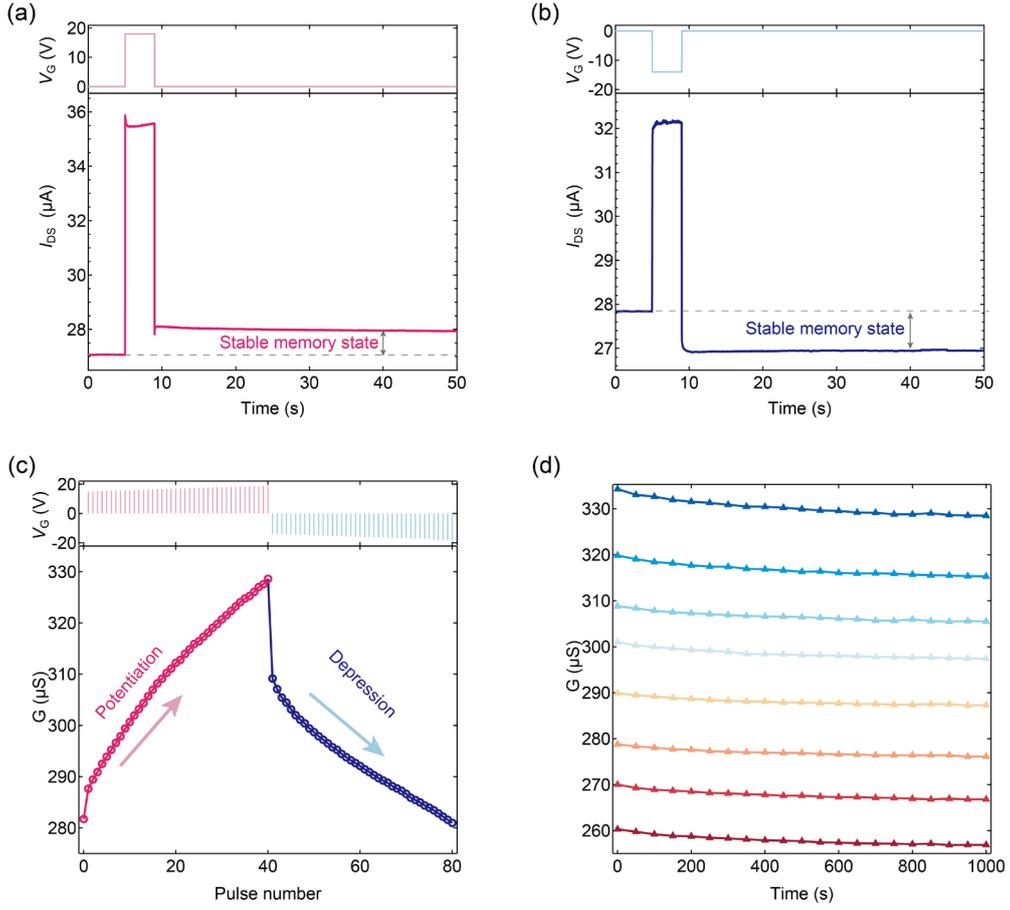

**Fig. 3. Characterization of long-term plasticity behavior of the moiré synaptic transistor**. (**a** and **b**) Current response induced by a larger positive $V_G$ pulse of 18 V for 4 s (a) or a larger negative $V_G$ pulse of -14 V for 4 s (b), exhibiting a stable memory state and emulating long-term potentiation behavior (a) or LTD behavior (b). (**c**) Conductance continuous modulation under 40 positive and 40 negative $V_G$ pulses with varying amplitude and fixed width of 4 s and interval of 10 s. The conductance values are read out at $V_{DS}$ = 0.1 V. (**d**) State retention of eight specific conductance states. Each state can be maintained for longer than 1000 s with low drift.

**Implementation of moiré physical neural network for reservoir computing**

Employing the highly tunable volatile and nonvolatile characteristics in our moiré synaptic transistor, we propose a full-moiré physical neural network (MPNN) for homogeneous-architecture hardware reservoir computing. The physical reservoir computing system consists of two parts: the reservoir layer based on the STP effect of the moiré transistors and responsible for nonlinearly extracting key features of input temporal information; and the readout layer implemented with the LTP characteristic of the device and responsible for feature analysis. For a proof-of-concept demonstration, we use the proposed MPNN framework to carry out the simulation of classification of the MNIST handwritten digit, as illustrated in Fig. 4a. The MPNN comprises of moiré synaptic transistor based 196 reservoir nodes and a 196×10 single-layer artificial neural network using the two moiré synaptic transistors as a differential pair. To feed the digit image into the network, the binarized image (28×28) is partitioned into seven columns and merged into a spatiotemporal 196×4 matrix. This matrix is then converted into 196 streams of 4-timeframe input pulses and is simultaneously applied to the gate terminal of the corresponding reservoir nodes. The resulting current at a specific moment, enabling the abstraction of various pattern features of input streams, is sampled from the reservoir node and recorded as the reservoir state. These reservoir states are subsequently fed into the readout layer for further analysis to perform training and inference for the classification task (see Supplementary materials for details).

Figure 4b demonstrates a typical encoding process for input patterns of "1100" and "0011" and the corresponding current response of the synaptic transistor, where each timeframe consists of a pulse of 18 V for 1 s, with the black pixel represented by constant voltage pulse and the white pixel represented by zero voltage pulse. For the same input patterns, we intentionally sample the resulting current responses at the moments of $t_1$ corresponding to the third timeframe and $t_2$ corresponding to the fourth timeframe, as the reservoir states, respectively. When both "1100" and "0011" are input to the device, we observe that the current response resulting from distinct patterns is different and distinguishable at $t_1$ and $t_2$, which can be attributed to the controllable fading memory characteristic of the device. Note that, the difference in read current is also influenced by the sampling time, since the current responses evolve over time. We apply 16 different input pulse streams, which cover all possible combinations of 4-bit patterns of transformed image matrices, to the device (Fig. S5), and sample the current states at $t_1$ and $t_2$, respectively. As demonstrated in Fig. 4c, the measured results from these two sampling modes exhibit distinguishable read current states, indicating that our moiré synaptic transistor can be used for reliable feature extraction of input spatiotemporal information as a moiré reservoir node. Notably, the two sets of sampled results give rise to distinct feature distributions, implying that different types of feature extraction can be implemented by using distinct sampling modes. These differences, arising from the stable nonlinear dynamics of the moiré synaptic transistor, enable the discrimination of diverse input patterns through current responses over an extended time scale. Combining the reservoir states recorded at various sampling moments allows for the fusion of various spatiotemporal features extracted, thereby fully exploiting the computational resources within the physical system. Based on these features, we further design a mixture sampling technique, in which each reservoir node selectively records the reservoir state at either $t_1$ or $t_2$ (Fig. S6). This unique sampling approach allows for fusing features extracted from different sampling modes without the need to increase the number of nodes. We employ these three sampling modes, *i.e.,* mixture sampling, sampling at $t_1$, and sampling at $t_2$, to process the handwritten digit images and acquire the corresponding reservoir states, which are subsequently projected to the readout layer for recognition. With such compressing of the spatial dimensional information, the input with the size of 784×10 is reduced to the size of 196×10. To train the weights in the readout layer, a total of 196×10 weight values corresponding to differential conductance states of the moiré synaptic device pair have been used (Fig. 4d).

We compare the training process of the reservoir states recorded by these three sampling modes (Fig. 4e). As the number of training epochs increases, the recognition accuracy of the test dataset improves, suggesting that the spatial features can be extracted through the reservoir layer. Notably, the stable and tunable dynamics of the device empower the mixture sampling mode to correlate and fuse multiple features. As a result, mixture sampling states can achieve a high recognition accuracy of 90.8% in a small-sized network, surpassing the 90.1% of the sampling states at $t_1$ and the 89.3% of the sampling states at $t_2$. Figure 4f further shows the inference results using mixture sampling states through a confusion matrix, wherein misclassifications predominantly occur among digits that are inherently challenging to distinguish, such as "3" and "8". These results clearly show that our proposed moiré synaptic transistor can be potentially used for implementing the homogeneous-architecture hardware of reservoir computing, and that diverse sampling modes can be utilized to selectively enhance the capability of feature extraction of various types of spatiotemporal information, which is inaccessible with previously reported devices.

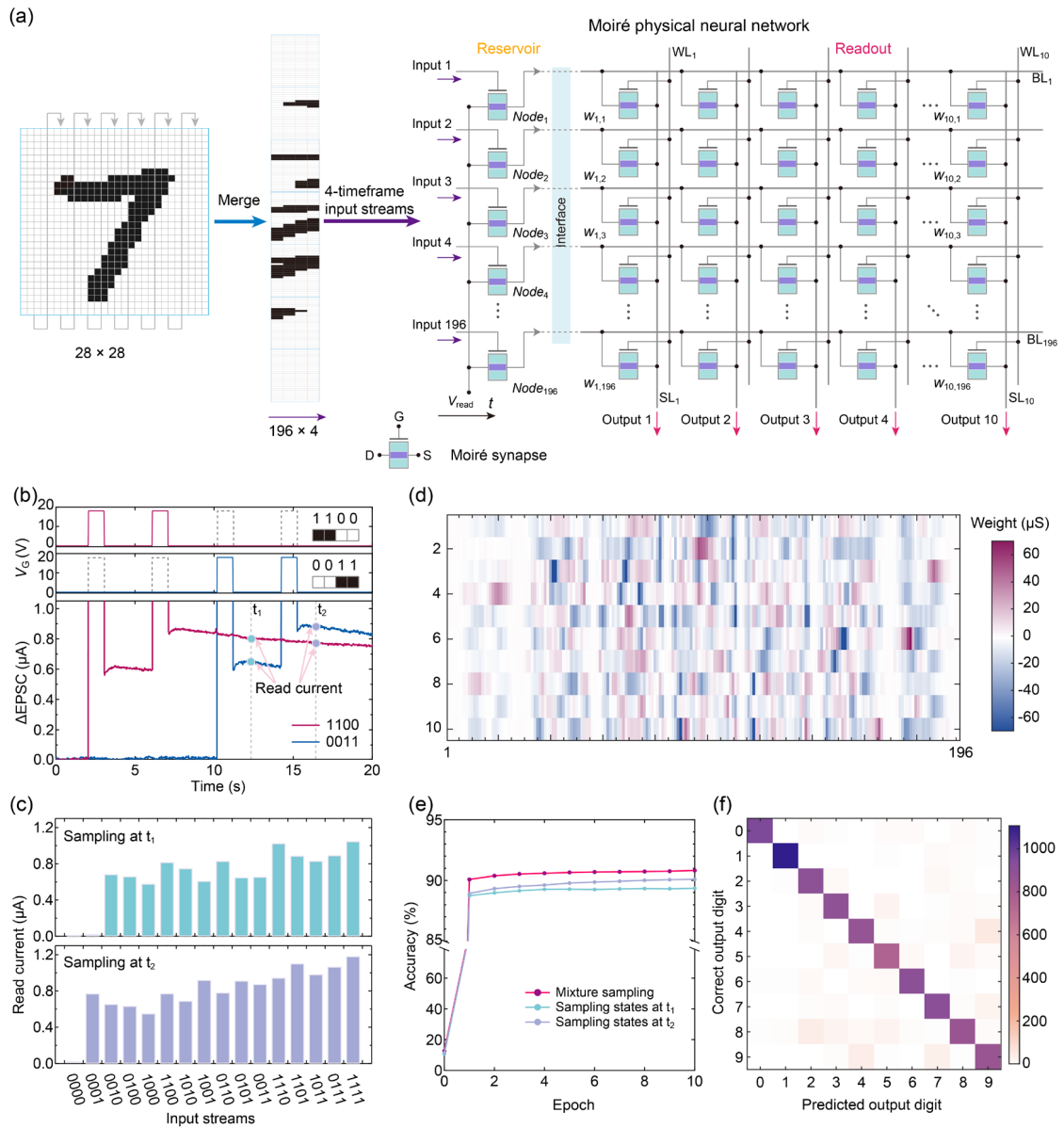

**Fig. 4. Implementation of moiré physical neural network for reservoir computing.** (a) Schematic illustration of the homogeneous architecture RC system based on MPNN to implement

handwritten digit recognition. Left, the binarized image with 28 × 28 pixels is divided into seven columns and merged into a 196×4 matrix, which is encoded as 196 4-timeframe input streams and sent into MPNN. Right, the MPNN consists of a reservoir layer involving 196 reservoir nodes, with each node corresponding to a moiré synaptic transistor, and a 196×10 readout network based on a 196×20 moiré synaptic transistor using differential pairs (differential pairs are not shown to simplify illustration). (**b**) Top, encoding process for "1100" and "0011" patterns. Bottom, current responses to 4-timeframe input streams of "1100" and "0011". The read voltage is fixed at 0.1 V. The moments $t_1$ and $t_2$ correspond to 1 second after the end of the pulse within the third and fourth timeframes, respectively. (**c**) Current states recorded at $t_1$ and $t_2$ of the moiré synapse device corresponding to all 4-timeframe input streams. (**d**) Final weight distribution of the readout layer after the 10 training epochs on the reservoir states obtained by mixture sampling. The weight can be expressed as $W=G^+-G^-$, where $G^{\pm}$ is the conductance of each device in the differential pair. (**e**) Simulated recognition accuracy of MNIST test dataset during 10 epochs. (**f**) Confusion matrix for classifying the test dataset.

**DISCUSSION**

In summary, we propose the moiré synaptic transistor based on the h-BN/bilayer graphene/h-BN moiré heterostructure and implement a homogeneous-architecture reservoir computing. The proposed synaptic transistor can be tuned to exhibit volatile and nonvolatile memory behaviors, faithfully emulating both short- and long-term synaptic plasticity in a single device. This moiré artificial synapse can serve as a fundamental building block for the reservoir and the readout layer, facilitating seamless integration within the reservoir computing system. Our work represents the initial effort in developing neuromorphic computing hardware based on the moiré quantum materials and opens up a promising avenue for the development of moiré neuromorphic computing. In the future, with controllable transfer techniques applied to large-area grown graphene and h-BN, it will be feasible to assemble a large-scale array of moiré heterostructures, thus promoting the hardware implementation of a full-moiré physical neural network.

**Acknowledgements**
This work was supported in part by the National Natural Science Foundation of China (62122036, 12322407, 62034004, 61921005, 12074176 and 61974176), the Strategic Priority Research Program of the Chinese Academy of Sciences (XDB44000000); the Fundamental Research Funds for the Central Universities (020414380203, 020414380179). F.M. acknowledges the support from the AIQ foundation.

# Supplementary Materials for

# "Moiré synaptic transistor for homogeneous-architecture reservoir computing"

Pengfei Wang(王鹏飞)[1][†], Moyu Chen(陈墨雨)[1][†], Yongqin Xie(谢永勤)[1], Chen Pan(潘晨)[2], Kenji Watanabe[3], Takashi Taniguchi[4], Bin Cheng(程斌)[2]*, Shi-Jun Liang(梁世军)[1]*, and Feng Miao(缪峰)[1]*

[1] Institute of Brain-Inspired Intelligence, National Laboratory of Solid State Microstructures, School of Physics, Collaborative Innovation Center of Advanced Microstructures, Nanjing University, Nanjing 210093, China.

[2] Institute of Interdisciplinary Physical Sciences, School of Science, Nanjing University of Science and Technology, Nanjing 210094, China.

[3] Research Center for Functional Materials, National Institute for Materials Science, 1-1 Namiki, Tsukuba 305-0044, Japan.

[4] International Center for Materials Nanoarchitectonics, National Institute for Materials Science, 1-1 Namiki, Tsukuba 305-0044, Japan.

† These authors contributed equally to this work.
*Corresponding authors. Email: bincheng@njust.edu.cn; sjliang@nju.edu.cn; miao@nju.edu.cn

**The PDF file includes:**
    Methods
    Figs. S1 to S6

## Methods

### Device fabrication

We fabricated a h-BN-encapsulated Bernal bilayer graphene (BLG) device with alignment between the straight edges of the BLG flake and the h-BN flakes by using a standard dry-transfer technique (Fig. S1). We first mechanically exfoliated bilayer graphene and h-BN (10-30nm) on 300-nm $SiO_2$/Si substrate, and selected the flakes of appropriate thickness and high quality with the assistance of optical contrast and atomic force microscopy (AFM). We used a poly (bisphenol A carbonate) (PC)/ polydimethylsiloxane (PDMS) stamp to stack a h-BN flake, a BLG flake, and another h-BN flake sequentially at 80°C. In the stacking process, the BLG sheet was first aligned with both the top and bottom h-BN flakes along the straight edges. The h-BN-encapsulated BLG heterostructure was then released from the PC film to a $SiO_2$/Si substrate at 140°C. The devices were etched by electron beam lithography and dry etching with a $CHF_3$/$O_2$ in an Inductively Coupled Plasma (ICP) system. The metal top gate and the edge contacts were patterned using the standard electron beam lithography method and deposited by the standard electron beam evaporation of Cr(5nm)/Pd(15nm)/Au(30nm).

### Electrical measurement

All the electrical measurements were performed at room temperature in a nitrogen atmosphere, using a semiconductor parameter analyzer (Keysight B1500A) and a probe station (Cascade Summit 11000 M). The input voltage pulses were programmed and generated through a high-voltage pulse generator unit (Keysight B1525A) or a source meter (Keithley 2636B). All measurements were controlled by the instruction implemented with LabVIEW.

### Implementation of handwritten digits recognition

To demonstrate the reliability of our proposed moiré physical neural network in implementing homogeneous architecture reservoir computing, we carried out a simulation for the recognition task of the MNIST handwritten digit dataset. The MNIST dataset consists of 60000 handwritten digits for training and another 10000 handwritten digits for testing. Each digit image in the dataset contains 28×28 pixels and has been binarized prior to classification. We first transformed the image into a 196×4 matrix, where each row corresponds to a 4-timeframe sequence. These sequences can be categorized into 16 different distributions, and all the 4-timeframe sequences in the matrix are a subset of these 16 possible sequences. To extract the spatial features of the 4-bit patterns processed by the moiré reservoir, we experimentally measured the read currents under 16 input pulse streams and obtained the reservoir states for each sequence. These measured current values were used to construct a 196×1 reservoir-processed feature vector for each image, where each element of the vector is represented by the read current extracted from the measured data for specific 4-bit input sequences. To achieve digit recognition based on the output results of the reservoir, we further used a fully connected 196×10 moiré-based network as the readout layer to process the 196×1 feature vector. We adopted the sigmoid activation function, which is defined as $f(x) = (1 + e^{-x})^{-1}$, and a typical cross-entropy loss function in network implementation. The neural network was trained by standard error backpropagation and mini-batch gradient descent algorithm with a batch size of 12 for weight updates. Considering the memory characteristics of the moiré synaptic transistor, a differential weight-mapping scheme was employed, where each effective synaptic weight contains a pair of moiré synaptic transistors, which is defined as $W = G^+ - G^-$. The conductance of the moiré synaptic transistor was experimentally demonstrated to be continuously tunable within the range of 260 μS to 330 μS, and the differential conductance adjust range was limited between -70 μS and 70 μS. In software, the corresponding weights were limited to the range of [-3.5, 3.5]. All the simulation algorithms were implemented using MATLAB.

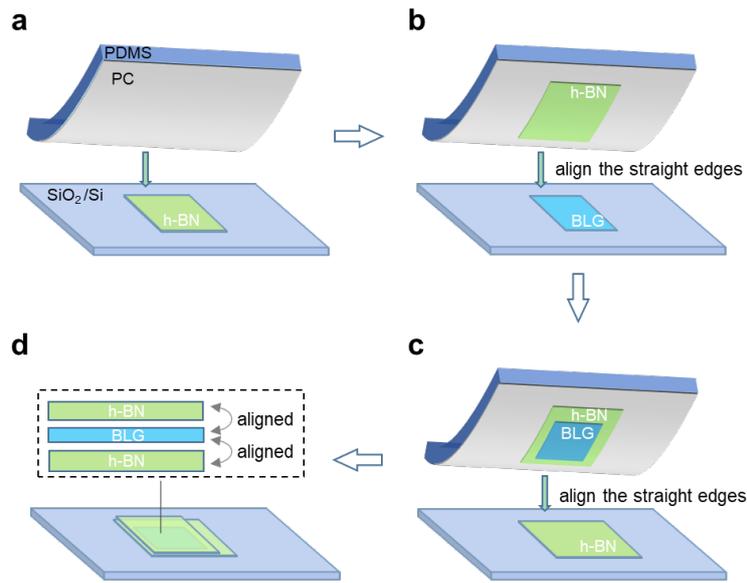

**Fig. S1. Dry transfer process for h-BN-encapsulated BLG heterostructure with alignment between the straight edges of BLG and h-BN.** (**a**) A h-BN flake with long straight edges was picked up by using a poly (bisphenol A carbonate) (PC)/ polydimethylsiloxane (PDMS) stamp at 80°C. (**b** and **c**) With the aid of a microscope, the straight edges of BLG and h-BN flakes can be precisely aligned. The BLG flake and another h-BN flake were stacked after alignment at 80 °C. (**d**) The h-BN-encapsulated BLG heterostructure was released from the PC film to a SiO$_2$/Si substrate at 140°C. The straight edge of the BLG flake was aligned with those of both the top and bottom h-BN flakes.

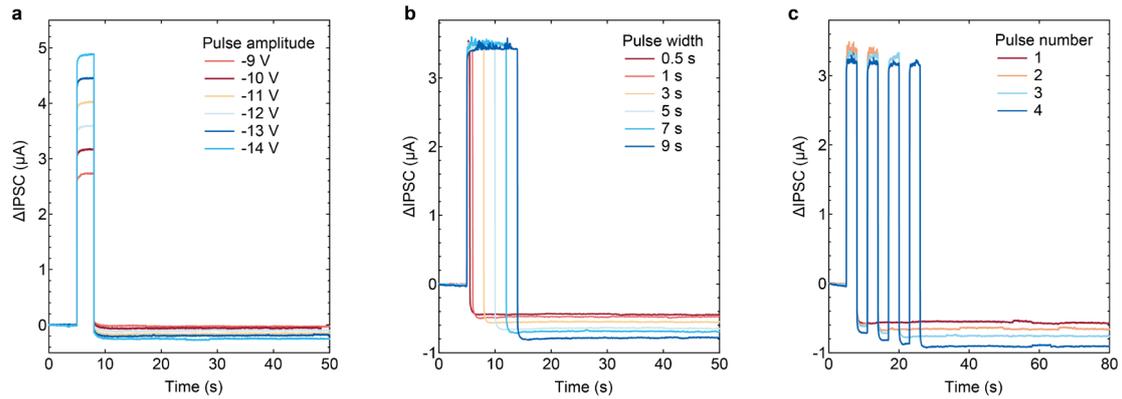

**Fig. S2. Dynamic current change of the moiré synaptic transistor in response to various external stimuli of negative $V_G$.** After removing the negative gate voltage pulse, the device current decreases below the initial state, mimicking inhibitory postsynaptic current (IPSC) behavior in biological synapses. (**a**) The IPSC responses to a train of pulse with different pulse amplitude from -9 V to -14 V and an identical width of 4 s. (**b**) The IPSC responses to a train of pulse with different pulse width from 0.5 s to 9 s but a fixed amplitude of -12 V. (**c**) The IPSC responses at different pulse number from 1 to 4, where the pulse amplitude, width, and interval are fixed at -12 V, 3 s, and 3 s, respectively. As the amplitude, width, and number of applied voltage pulses increase, the current change exhibits a gradual increase, indicating a transition from short-term depression to complete long-term depression.

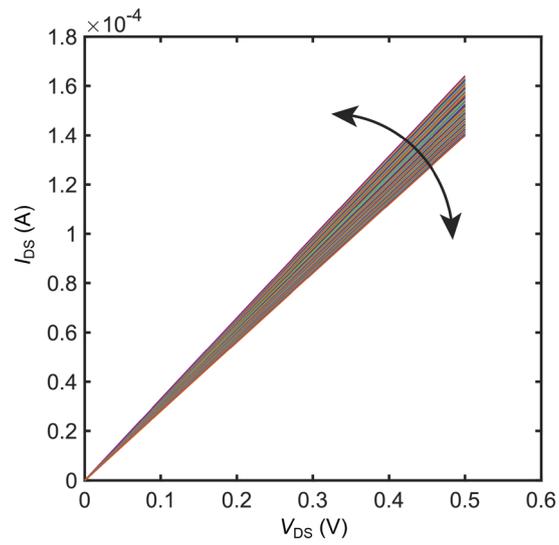

**Fig. S3. Linear current-voltage characteristics of the moiré synaptic transistor with different conductance states.** Under the precise control of the gate voltage pulse, the conductance of the moiré synaptic transistor can be continuously tuned within a specific range. The device demonstrates excellent linear *I-V* characteristics, making it a suitable synaptic unit for the artificial neural network hardware, which can be used to physically perform multiplication through Ohm's law: $I=G \cdot V$.

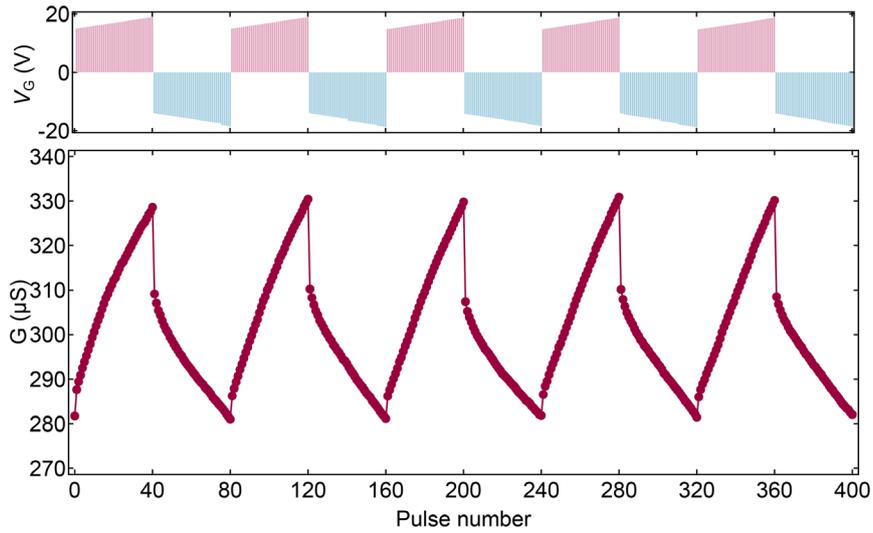

**Fig. S4. Reproducibility of long-term potentiation and depression at a moiré synaptic transistor.** Top, the device was programmed by a sequence of 40 positive pulses (amplitude: from 15 to 19 V in increments of 0.1V; width: 4 s) followed by 40 negative pulses (amplitude: from -14 to -18 V in increments of -0.1V; width: 4 s). Bottom, the conductance incrementally changes through this sequence of pulses, exhibiting long-term potentiation and depression characteristics. This sequence was repeated five times, and the device demonstrates great repeatability and negligible cycle-to-cycle variation. The conductance values were read out at $V_{DS}$ = 0.1 V after each pulse.

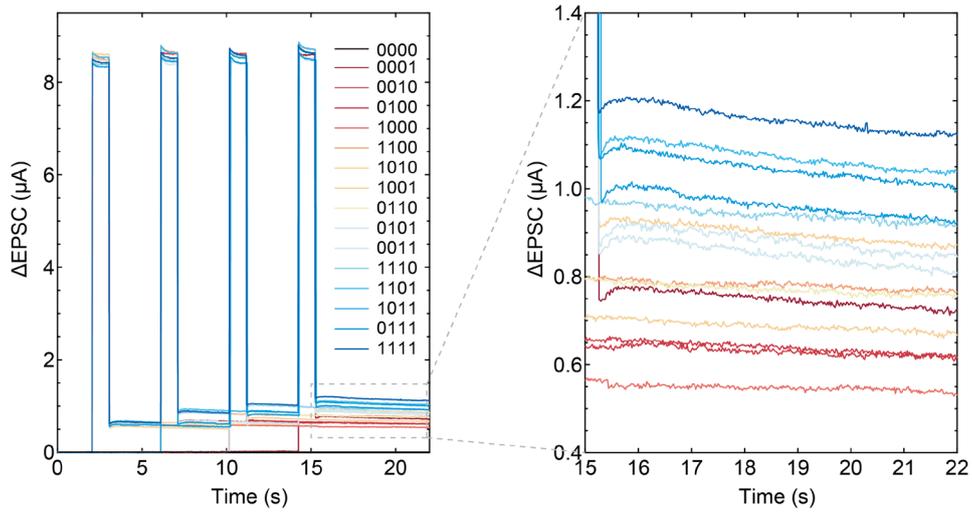

**Fig. S5. Dynamic current change in response to 16 different input pulse streams ranging from "0000" to "1111".** The bit "1" is represented by a high amplitude pulse (18 V, 1 s), and the bit "0" is represented by a pulse with an amplitude of 0 V (equivalently no pulse). Right, plot of the enlargement for the gray dashed box. Due to the short-term memory effect and nonlinearity of the device, the current response to 16 different input pulse streams can be clearly distinguishable.

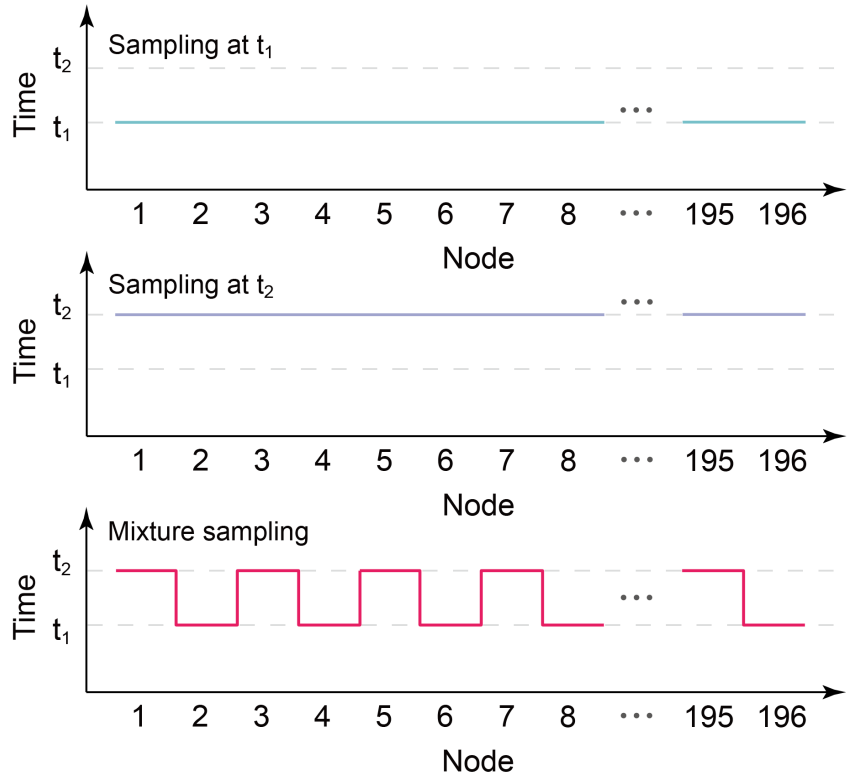

**Fig. S6. Three reservoir state sampling modes.** The reservoir layer comprises 196 reservoir nodes, where the encoded 196 streams of 4-timeframe pulses are input into the corresponding reservoir nodes respectively. To record the reservoir states, we categorize the sampling modes into three types according to the sampling time for each node: all nodes sampling at $t_1$, all nodes sampling at $t_2$, and mixture sampling. In the mixture sampling mode, the sampling time alternates between $t_1$ and $t_2$ for each node.